\pdfoutput=1 
\documentclass[prl,twocolumn,showpacs,superscriptaddress]{revtex4}

\usepackage{amsmath,amssymb}
\usepackage{verbatim}
\usepackage{graphicx}
\usepackage{color}
\usepackage[colorlinks=true,linkcolor=red,citecolor=magenta,urlcolor=blue,filecolor=black]{hyperref}

\DeclareFontFamily{OT1}{rsfs}{}
\DeclareFontShape{OT1}{rsfs}{m}{n}{ <-7> rsfs5 <7-10> rsfs7 <10->rsfs10}{} 
\DeclareMathAlphabet{\mycal}{OT1}{rsfs}{m}{n}

\newcommand{\be}[1]{ \begin{equation}\label{#1} }
\newcommand{\ee}{\end{equation}}
\newcommand{\bea}[1]{\begin{eqnarray}\label{#1} }
\newcommand{\eea}{\end{eqnarray}}

\DeclareMathOperator{\extdm}{d}
\newcommand{\extd}{\extdm \!}

\begin{document}


\title{Uniformization of entanglement entropy in holographic warped conformal field theories}



\author{St\'ephane Detournay}
\email{sdetourn@ulb.ac.be}
\affiliation{Physique Th{\'e}orique et Math{\'e}matique, Universit{\'e} libre de Bruxelles and International Solvay Institutes, Campus Plaine C.P. 231, B-1050 Bruxelles, Belgium}

\author{Daniel Grumiller}
\email{grumil@hep.itp.tuwien.ac.at}
\affiliation{Institute for Theoretical Physics, TU Wien, Wiedner Hauptstrasse 8--10/136, A-1040 Vienna, Austria}

\author{Max Riegler}
\email{mriegler@fas.harvard.edu}
\affiliation{Center for the Fundamental Laws of Nature, Harvard University, Cambridge, MA 02138, USA}

\author{Quentin Vandermiers}
\email{qvdmiers@ulb.ac.be}
\affiliation{Physique Th{\'e}orique et Math{\'e}matique, Universit{\'e} libre de Bruxelles and International Solvay Institutes, Campus Plaine C.P. 231, B-1050 Bruxelles, Belgium}
\affiliation{Institute for Theoretical Physics, TU Wien, Wiedner Hauptstrasse 8--10/136, A-1040 Vienna, Austria}

\date{\today}

\preprint{TUW--20--03}

\begin{abstract}

Using a uniformization map we determine the holographic entanglement entropy for states of a Warped Conformal Field Theory dual to a generic vacuum metric in AdS$_3$ gravity with Comp\`ere--Song--Strominger boundary conditions. We point out how that expression could lead to inequalities that can be interpreted as quantum energy conditions for Warped Conformal Field Theories.



\end{abstract}


\maketitle

\section{Introduction and summary}
The stress-tensor $T_{\mu \nu}$ of a Quantum Field Theory (QFT) dictates through Einstein's equations contraints on the geometry arising semi-classically when coupling gravity to matter described by this QFT. Various energy conditions on $T_{\mu \nu}$ can be formulated, expressing for instance the positivity of energy density (Weak Energy Condition) or causal propagation of energy flow (Dominant Energy Condition). A weaker energy condition is the Null Energy Condition (NEC).
\begin{equation}
  T_{\mu \nu} k^\mu k^\nu \geq 0 \qquad \forall k^\mu\;  | \;  k^\mu k_\mu = 0
\end{equation}
The proofs of the black hole area law \cite{Hawking:1971tu} or singularity theorems \cite{Penrose:1964wq} crucially rely on the NEC. This condition, however, is violated quantum-mechanically, e.g.~in the Casimir effect or by Hawking radiation. Instead, quantum mechanically QFTs typically obey non-local conditions such as the Averaged NEC (see e.g.~\cite{Faulkner:2016mzt,Hartman:2016lgu} for recent proofs and refs.~therein), which states that negative energy fluxes along null directions are compensated by positive energy fluxes (with ``quantum interest'' \cite{Ford:1994bj}). 

The Quantum Null Energy Condition (QNEC) \cite{Bousso:2015mna} is a local energy condition conjectured to extend NEC to the quantum regime, and has attracted a lot of attention in recent years \cite{Fu:2016avb,Fu:2017evt,Akers:2017ttv,Ecker:2017jdw,Fu:2017ifb,Leichenauer:2018obf,Khandker:2018xls,Ecker:2019ocp}, including proofs for free QFTs \cite{Bousso:2015wca}, for holographic Conformal Field Theories (CFTs) \cite{Koeller:2015qmn}, then for general CFTs \cite{Balakrishnan:2017bjg}, and shown to hold universally for generic QFTs under the same assumptions required for the averaged NEC \cite{Ceyhan:2018zfg}. For two-dimensional CFTs (CFT$_2$), QNEC reads \cite{Bousso:2015mna,Wall:2011kb}
\begin{equation}
	2 \pi \left< T_{\mu\nu} k^\mu k^\nu\right> \geq S'' + \frac{6}{c}\, S^{\prime\,2} \qquad \forall k^\mu\;  | \;  k^\mu k_\mu = 0 \label{QNECCFT}
\end{equation}
where $c$ is the central charge of the CFT, $\left< T_{\mu\nu}k^\mu k^\nu\right>$ the expectation values of the null projections of the stress tensor for a given state and $S$ is the entanglement entropy (EE) for an arbitrary interval of this state; prime denotes variations of EE with respect to null deformations in the null direction defined by $k^\mu$ of one of the endpoints of the entangling region. 

In the context of AdS$_3$/CFT$_2$, it was shown that QNEC saturates not only for the vacuum, for states dual to particles on AdS$_3$ or BTZ black holes, or for any state that is a Virasoro descendant thereof \cite{Khandker:2018xls}, but also for all states dual to Ba\~nados geometries \cite{Banados:1998gg}, some of which describe systems far from thermal equilibrium \cite{Ecker:2019ocp}. This was done exploiting the fact that all Ba\~nados geometries are locally AdS$_3$ and using a uniformization map between Poincar\'e AdS$_3$ and the Ba\~nados geometries \cite{Sheikh-Jabbari:2016unm}.

Following a similar strategy, a Quantum Energy Condition was derived recently \cite{Grumiller:2019xna} for a class of non-Lorentz invariant holographic theories with BMS$_3$ symmetries, through a uniformization map between Minkowski space and the flat version of Ba\~nados geometries \cite{Barnich:2012aw}, yielding inequalities involving the supertranslation and superrotation fields instead of the CFT stress-tensor.

In this note, we take the first steps towards extending these results to another class of non-relativistic theories, Warped Conformal Field Theories (WCFTs) \cite{Detournay:2012pc}. We first review the results for AdS$_3$ gravity with Brown--Henneaux boundary conditions and the derivation of the satured version of \eqref{QNECCFT} for Ba\~nados geometries. We then turn to a simple holographic model for WCFTs, consisting in pure Einstein gravity in $2+1$ dimensions with a negative cosmological constant and chiral/Comp\`ere--Song--Strominger (CSS) boundary conditions \cite{Compere:2013bya}. The role of the Ba\~nados geometries there is played by a gauge-fixed and on-shell version of the CSS boundary conditions, referred to as CSS geometries. We determine a uniformization map that allows us to derive EE for states of a WCFT dual to these geometries. We express components of the holographic stress tensor in a form reminiscent of the saturated form of QNEC \eqref{QNECCFT}. Whether these can be turned into Quantum Energy Conditions for WCFTs will be explored in a forthcoming work \cite{inprogress}. 


\section{Warped Conformal Field Theories}
A WCFT is a two-dimensional QFT invariant under ``warped conformal symmetries''. Parametrizing the plane with coordinates $t_\pm$, these are given by
\begin{equation}
  t_+ \rightarrow f(t_+) \qquad \qquad  t_- \rightarrow t_- +  g(t_+) \label{WCT}
\end{equation}
where $f(t_+)$ and $g(t_+)$ are two arbitrary functions. 
These theories are not Lorentz-invariant, and the symmetries (\ref{WCT}) can be shown to arise by assuming translation invariance and chiral scaling \cite{Hofman:2011zj}, comparable to the emergence of local conformal symmetry in unitary Poincar\'e-invariant two-dimensional QFTs with a global scaling symmetry and a discrete non-negative spectrum of scaling dimensions \cite{Polchinski:1987dy}. Reparametrizations of $t_+$ and coordinate-dependent translations of $t_-$ are generated by a stress tensor $T(t_+)$ and current density $P(t_+)$, whose modes span a Virasoro-Ka\v{c}-Moody algebra with global $\mathfrak{sl}(2,R) \oplus \mathfrak{u}(1)$ subalgebra. 

The study of WCFTs was triggered by the search for holographic duals to the near-horizon region of extremal black holes \cite{Bardeen:1999px, Dias:2007nj, Guica:2008mu} and Warped AdS$_3$ (WAdS$_3$) spaces \cite{Banados:2005da, Anninos:2008fx}, exhibiting an $SL(2,R) \times U(1)$ symmetry. The asymptotic symmetry algebra (ASA) of WAdS$_3$ spaces was shown to precisely consist in a Virasoro--Ka\v{c}--Moody algebra \cite{Compere:2007in, Compere:2008cv, Compere:2009zj,Blagojevic:2009ek, Henneaux:2011hv}. 

The holographic duality relating WAdS$_3$ spaces and WCFTs has passed several tests, including the matching of Bekenstein--Hawking entropy \cite{Detournay:2012pc}, of greybody factors from correlation functions \cite{Song:2017czq} and of one-loop determinants in the bulk from characters \cite{Castro:2017mfj}. Explicit examples of WCFTs have appeared in \cite{Compere:2013bya, Hofman:2014loa, Castro:2015uaa, Jensen:2017tnb}. Importantly for this work, universal expressions for EE in WCFTs were derived holographically \cite{Castro:2015csg, Song:2016pwx, Song:2016gtd}. 



\section{Saturated QNEC for hologaphic CFT$_2$}

In this section we review a holographic derivation of the saturated QNEC for CFT$_2$. In AdS$_3$ gravity with Brown--Henneaux boundary conditions, the most general vacuum solution in Fefferman-Graham gauge is the Ba\~nados metric \cite{Banados:1998gg} (see also \cite{Sheikh-Jabbari:2016unm, Compere:2015knw})
\begin{multline}
	\frac{\extd s^2}{\ell^2} = \frac{\extd z^2 - \extd x_+ \extd x_-}{z^2} + L_+ \extd x_+^2 + L_- \extd x_-^2 \\
	- z^2 L_+ L_- \extd x_+ \extd x_- \label{Banados metric}
\end{multline}
where $L_\pm = L_\pm (x_\pm)$ and $\ell$ is the AdS radius.

 The expectation values of the stress tensor are related to the functions $L_\pm$ present in the Ba\~nados metric \eqref{Banados metric} by
\begin{equation}
	2\pi \left< T_{\pm \pm} \right> = \frac{c}{6} L_{\pm} \label{relation L T}
\end{equation}
where $c$ is the Brown--Henneaux central charge \cite{Brown:1986nw}. The Poincar\'e patch is just a special case where $L_\pm =0$ in \eqref{Banados metric}. As the Ba\~nados metric is locally AdS$_3$, there is a mapping for the Poincar\'e patch to \eqref{Banados metric} \cite{Roberts:2012aq, Sheikh-Jabbari:2016znt}:
\begin{subequations}
\label{Banados diff}
\begin{align}
	x^\pm_P &= \int \frac{dx_\pm}{\psi^{\pm 2}} - \frac{z^2 \psi^{\mp \prime}  }{\psi^\pm \psi^\mp (1-z^2/z_h^2)} \\
	z_P &= \frac{z}{\psi^+ \psi^- (1-z^2/z_h^2)} 
\end{align}
\end{subequations}
where the functions $\psi^\pm(x_\pm)$ obey Hill's equation:
\begin{equation}
	\psi^{\pm \prime \prime} - L_\pm \psi^\pm = 0 \label{Hill eq}
\end{equation}
and $z_h$ is one of the Killing horizons of the Ba\~nados metric \eqref{Banados metric}. Expressing the two independent solutions of Hill's equation by $\psi_{1,2}^\pm$, it is convenient to normalize them as 
\begin{equation}
	\psi^\pm_1 \psi^{\pm \prime}_2 - \psi^{\pm \prime}_1 \psi^\pm_2 = \pm 1\,.
\end{equation}

From this diffeomorphism, we can find the holographic EE of the Ba\~nados metric starting from the one of the Poincar\'e patch \cite{Ryu:2006bv}:
\begin{equation}
	S_{PP} = \frac{c}{3} \ln\frac{l}{\epsilon} = \frac{c}{6} \ln\left(\frac{(x_1^+-x_2^+) (x_1^- - x_2^-)}{\epsilon^2} \right) \label{EE Poincare}
\end{equation}
where $x_{1,2}^\pm$ are the boundary points of the entangling interval $l$. It is sufficient to know the near boundary behaviour of \eqref{Banados diff} to compute the holographic EE. Close to the boundary, one has the conformal transformation
\begin{equation}
	x_P^\pm = \frac{\psi_1^\pm}{\psi_2^\pm} \qquad z_p = \frac{z}{\psi_2^+ \psi_2^-}	
\end{equation}
and one finds \cite{Sheikh-Jabbari:2016znt}
\begin{equation}
	S_{\textrm{\tiny{HEE}}} =\frac{c}{6} \ln \frac{l^+(x_1^+,x^+_2) l^-(x_1^-,x_2^-)}{\epsilon^2} =: S_+ + S_- \label{EE CFT}
\end{equation}
with
\begin{equation}
	l^\pm(x_1^\pm,x^\pm_2) = \psi_1^\pm(x_1^\pm) \psi_2^\pm(x_2^\pm) - \psi_1^\pm(x_2^\pm) \psi_2^\pm(x_1^\pm)\,. \label{l def}
\end{equation}
We are now in position to show that the Ba\~nados geometries saturate QNEC \cite{Ecker:2019ocp}. Defining the `vertex function'
\begin{equation}
	V :=\exp\Big(\frac{6}{c} S\Big) = \frac{l^+(x_1^+,x^+_2)\, l^-(x_1^-,x_2^-)}{\epsilon^2} \label{V def}
\end{equation}
it is straightforward to show that it satisfies Hill's equation \eqref{Hill eq},
\begin{equation}
	V'' = L_\pm V\,.
\end{equation}
On the other hand, the definition \eqref{V def} implies
\begin{equation}
	\frac{V''}{V} = \frac{6}{c}\, \Big(S'' + \frac{6}{c} S^{\prime 2} \Big)\,.
\end{equation}
Now, using the relation between the stress tensor and the functions $L_\pm$, proves that for the Ba\~nados metric the QNEC inequality \eqref{QNECCFT} saturates.
\begin{equation}
	2 \pi \left< T_{\pm \pm}\right> = S_\pm'' + \frac{6}{c}\, S_\pm'^2
\label{eq:angelinajolie}
\end{equation}

\section{Holographic WCFT model}

%
%


We take as our holographic model AdS$_3$ gravity (again with AdS radius $\ell$) with CSS boundary conditions \cite{Compere:2013bya}. The counterpart of the Ba\~nados metric is given by 
\begin{align}
	\frac{\extd s^2}{\ell^2} &= \frac{\extd z^2}{z^2} - \Big( \frac{1}{z^2} + \frac{2 \Delta}{k}  P' + \frac{\Delta}{k^2} L  z^2\Big) \extd t_+ \extd t_-  + \frac{\Delta}{k}  \extd t_-^2  \nonumber \\
	&+ \Big( \frac{P'}{z^2} + \frac{1}{k} (L + \Delta \ P'^2) + \frac{\Delta}{k^2} L \ P' \ z^2 \Big) \extd t_+^2 \label{CSS metric}
\end{align}
where $k = \ell/4G$, $\Delta$ is a constant and the functions $P' =: \partial_+ P$ and $L$ depend on $t_+$ only. If the latter functions vanish, we recover an extremal BTZ black hole with $\ell M - J = 0$ and $\ell M  + J = \Delta$ \cite{Compere:2013bya}. The asymptotic boundary is at $z=0$; to map it to $r=+\infty$ we use from now on $r=1/z$ as radial coordinate.

Asymptotically the CSS metric \eqref{CSS metric} has a Virasoro--Ka\v{c}--Moody algebra symmetry that acts at the boundary as
\begin{equation}
	\begin{array}{l}
		t_+ \rightarrow f(t_+) \\
		t_-  \rightarrow t_- + g(t_+)
	\end{array} \label{WCFT transf}
\end{equation}
and is generated infinitesimally by the asymptotic Killing vectors
\begin{equation}
	\begin{array}{l}		
		\xi (\epsilon) =  \epsilon(t_+) \partial_+ -\frac{r}{2} \epsilon'(t_+) \partial_r +\text{subleading} \\
		\eta (\sigma) =  \sigma(t_+) \partial_- + \text{subleading}
	\end{array}\label{AKV}
\end{equation}
The corresponding charges generating the Virasoro--Ka\v{c}--Moody algebra are given by 
\begin{subequations}
 \label{Charges}
\begin{align}
  Q_\epsilon &= \frac{1}{2\pi} \int \extd \phi \ \epsilon (t_+) \left( L - \Delta  P'^2 \right)\\ 
  Q_\sigma &= \frac{1}{2\pi} \int \extd \phi \ \sigma (t_+) \left( \Delta + 2 \Delta P' \right)
  \end{align}
  \end{subequations}

\section{Uniformized warped entanglement entropy}

 In this section  we derive the EE expressions for the family of metrics \eqref{CSS metric} after deriving a warped version of the uniformization procedure reviewed above.

\subsection{Subleading terms}

We first derive the explicit form of the infinitesimal diffeomorphism (\ref{AKV}).
The infinitesimal transformations leaving (\ref{CSS metric}) invariant are of the form \cite{Barnich:2012aw, Compere:2015knw}:
\begin{equation}
	\chi^r = r \ \sigma(r) \qquad \chi^a= \epsilon^a(x^b)- \ell^2 \partial_b \sigma \ \int\limits_r^\infty \frac{\extd r'}{r'} \gamma^{ab}(r',x^a) \label{inf dif chi}
\end{equation}
where $\extd s^2 = \ell^2 \frac{\extd r^2}{r^2} + \gamma_{ab}(r,x^c) \extd x^a \extd x^b$, $\epsilon$ is a conformal Killing vector at the $r=\infty$ boundary and $\sigma$ is the Weyl factor of $\epsilon$. 


The explicit results
\begin{subequations}
\label{chi}
\begin{align}
	\chi^r &= - \frac{r}{2} \epsilon' \\
	\chi^+ &= \epsilon(t_+) + \frac{k \Delta \epsilon''}{2(k^2 r^4 - L \Delta)}  \\
	\chi^- &=  \sigma(t_+) + \frac{k  (k r^2 + P' \Delta)\epsilon''}{2 (k^2 r^4 - L \Delta)}
\end{align}
\end{subequations}
yield finite variations of the functions defining the physical state
\begin{eqnarray}
	\delta_{\chi} L &=& 2 \epsilon' L + \epsilon L' - \frac{k}{2} \epsilon''' \\
	\delta_{\chi} P' &=&(\epsilon P'  - \sigma)'
\end{eqnarray}
We recover the same infinitesimal transformation for $L$ as in the Ba\~nados metric. This is expected, since in both cases there is an underlying Virasoro symmetry. The transformation of $P'$, however, is not governed by Virasoro symmetries; instead, it is governed by $\mathfrak{u}(1)$ Ka\v{c}--Moody symmetries.


\subsection{Uniformization from extremal BTZ}

In the boundary conditions of \cite{Compere:2013bya}, $\Delta$ is a fixed constant, and one cannot reach a metric \eqref{CSS metric} with $\Delta \ne 0$ from one with $\Delta = 0$, in particular Poincar\'e AdS$_3$.\footnote{This is related to the fact that the CSS boundary conditions are dual to a WCFT in the so-called quadratic ensemble, in which the level is $U(1)$ charge-dependent, the latter not being able to vary over phase space -- here the zero mode $U(1)$ charge is given by $\Delta$. There exist alternative boundary conditions yielding a constant level and a varying zero mode $U(1)$ charge, see Appendix of \cite{Compere:2013bya} or the AdS$_3$ limit of WAdS$_3$ boundary conditions of \cite{Compere:2009zj}. These are naturally dual to a WCFT in the canonical ensemble. See e.g.~\cite{Apolo:2018eky} for a discussion on the relation between the two ensembles.} 
Therefore, let us consider a CSS metric with $\Delta \ne 0$ and vanishing $P'$ and $L$.
\begin{equation}
	\frac{\extd s^2_P}{\ell^2} = \frac{\extd u^2 - \extd y_+ \extd y_-}{u^2} \ + \frac{\Delta}{k} \extd y_-^2 \label{Poincare metric}
\end{equation}
The change of coordinates between (\ref{Poincare metric}) and (\ref{CSS metric}) is given by 
\begin{eqnarray}
	y_+ &=& \int \frac{\extd t_+}{\psi^2} - \frac{ \Delta \psi' z^4}{\psi( k \psi^2 - \Delta \psi^{\prime 2} z^4) } \nonumber \\
	y_- &=& t_- - C(t_+) - \sqrt{\frac{k}{\Delta}} \text{artanh} \left(\sqrt{\frac{\Delta}{k}}  \frac{\psi^{\prime}}{\psi} z^2  \right)   \label{dif BTZ to CSS} \\
	u &=& \frac{\sqrt{k}z}{\sqrt{k \psi^2 - \Delta \psi^{\prime 2} z^4}} \nonumber 
\end{eqnarray}
where $C(t_+)$ and $\psi(t_+)$ are given by the warped analogue of Hill's equation.
\begin{equation}
	P'(t_+) = C'(t_+) \qquad \psi'' - \frac{L(t_+)}{k} \psi = 0  \label{Hill eq2}
\end{equation}

\subsection{Uniformized entanglement entropy}

EE for a WCFT in a state dual to \eqref{Poincare metric} is given by \cite{Castro:2015csg, Song:2016pwx, Song:2016gtd, Azeyanagi:2018har}\footnote{To recover the metric \eqref{Poincare metric} from the section 5 in \cite{Azeyanagi:2018har}, one has to choose the parameters as $\alpha = 4/\Delta$, $b=1/\sqrt{2}$, $c=2$, makes the change of radial coordinate $e^{-\rho} = \sqrt{\Delta/k} z^2$ and sets the functions $\mathfrak{L}$ and $\mathcal{K}$ to zero.}
\begin{equation}
	S_{\textrm{\tiny{EE}}} = - \sqrt{\Delta k} (y_1^- - y_2^-) + k \ln \left[\sqrt{\frac{k}{\Delta}} \frac{y_1^+ - y_2^+}{2 \epsilon^2} \right] \label{EE Delta}
\end{equation}
where $y_i^\pm$ are the endpoints of the interval and $\epsilon$ a UV cut-off.

Performing the diffeomorphism \eqref{dif BTZ to CSS} yields the EE of the WCFT state dual to the CSS metric
\begin{equation}
	S_{\textrm{\tiny{EE}}} = S_P + S_L \label{EE CSS}
\end{equation}
where we separated the entropy in a Ka\v{c}-Moody part $S_P$ and a Virasoro part $S_L$
\begin{align}
	S_P &=  - \sqrt{\Delta k} (t_1^- - t_2^--C(t_1^+)+C(t_2^+))  \\
	S_L &= \frac{c}{6} \ln \left[\sqrt{\frac{k}{\Delta}} \frac{l^+(t_1^+,t_2^+)}{2 \epsilon^2}\right]
\end{align}
and used the same normalization for $\psi$ than in the CFT case and \eqref{l def}.

\subsection{WCFT saturation equations}

The AdS$_3$ stress tensor \cite{Henningson:1998gx, Balasubramanian:1999re, Kraus:2005zm, Kraus:2006wn} for the CSS boundary conditions reads~\footnote{As argued in \cite{Hofman:2014loa}, the background geometry to which WCFTs couple is not Riemannian, but rather has Newton--Cartan structure. From this perspective, the so-defined stress tensor is not the most natural object to consider, but for now we take \eqref{Tmunu} as a useful way to repackage the WCFT currents generating the conserved charges.}
\begin{equation}
	\frac{12 \pi}{c} \left<  T_{ab}\right> = g_{ab}^{(2)} - g^{kl}_{(0)}g_{kl}^{(2)} g_{ab}^{(0)}	
	 = \begin{pmatrix}
	\frac{L}{k} + \frac{\Delta}{k} P'^2 & -\frac{\Delta}{k} P' \\
	-\frac{\Delta}{k} P' & \frac{\Delta}{k}
	\end{pmatrix} \,.\label{Tmunu}
\end{equation}
Using (\ref{EE CSS}), its components are shown to satisfy 
\begin{eqnarray}
	2 \pi \left<  T_{++}\right> &=& S_L'' + \frac{6}{c} (S_L'^2 + S_P'^2) \\
	2 \pi \left<  T_{+-}\right> &=& \frac{6}{c} S_P' \dot{S}_P \\
	2 \pi \left<  T_{--}\right> &=& \frac{6}{c} \dot{S}_P^2 \label{QEC --}
\end{eqnarray}
where \textit{prime} denotes a derivation with respect to $t_+$ and the \textit{dot} a derivation with respect to $t_-$.
Another set of relations can be derived in terms of the currents responsible for the Virasoro-Ka\v{c}-Moody charges (\ref{Charges}). Defining
\begin{equation}
2 \pi \left<  T_L\right> = L - \Delta P'^2 \quad , \quad 2 \pi \left< T_P\right> = \Delta + 2 \Delta P'
\end{equation}
one has 
\begin{align}
2 \pi \left< T_L\right>  &= S_L'' + \frac{6}{c} ( S_L'^2 - S_P'^2) \\
2 \pi \left< T_P\right> &=  \frac{6}{c}( \dot{S}_P^2 - 2  S_P'   \dot{S}_P)\,.
\end{align}
These equalities are WCFT analogues of the QNEC saturation equations \eqref{eq:angelinajolie}. Whether or not some of them can be turned into inequalities when matter is added will be reported elsewhere \cite{inprogress}.
\newline
\newline

\acknowledgments

We thank Georg Stettinger and Raphaela Wutte for discussions and collaboration on related topics \cite{inprogress}.
SD is a Research Associate of the Fonds de la Recherche Scientifique F.R.S.-FNRS (Belgium). SD was supported in part by IISN -- Belgium (convention 4.4503.15) and benefited from the support of the Solvay Family. SD acknowledges support of the Fonds de la Recherche Scientifique F.R.S.-FNRS (Belgium) through the CDR project C 60/5 - CDR/OL ``Horizon holography : black holes and field theories" (2020-2022). 
DG is supported by the Austrian Science Fund (FWF), projects P~30822 and P~32581. 
The research of MR is supported by the European UnionÕs Horizon 2020 research and innovation programme under the Marie Sk?odowska-Curie grant agreement No 832542.
QV thanks DG for inviting him at the TU Wien and the hospitality of the Vienna group during his stay.


\bibliographystyle{apsrev}
\bibliography{Bibliography}

\end{document}